\documentclass[aps,pra,superscriptaddress,twocolumn,showpacs]{revtex4}
\usepackage{graphicx,psfrag}

\begin{document}

\title{Nonclassicality in Weak Measurements}

\author{Lars M. Johansen}
\affiliation{Department of Technology, Buskerud University College,
N-3601 Kongsberg, Norway} \email{lars.m.johansen@hibu.no}

\author{Alfredo Luis}
\affiliation{Departamento de \'{O}ptica, Facultad de Ciencias
F\'{i}sicas, Universidad Complutense, 28040 Madrid, Spain}
\email{alluis@fis.ucm.es}

\date{\today}

\begin{abstract}

We examine weak measurements of arbitrary observables where the
object is prepared in a mixed state and on which measurements with
imperfect detectors are made. The weak value of an observable can be
expressed as a conditional expectation value over an infinite class
of different generalized Kirkwood quasi-probability distributions.
``Strange" weak values for which the real part exceeds the
eigenvalue spectrum of the observable can only be found if the
Terletsky-Margenau-Hill distribution is negative, or, equivalently,
if the real part of the weak value of the density operator is
negative. We find that a classical model of a weak measurement
exists whenever the Terletsky-Margenau-Hill representation of the
observable equals the classical representation of the observable and
the Terletsky-Margenau-Hill distribution is nonnegative. Strange
weak values alone are not sufficient to obtain a contradiction with
classical models.

We propose feasible weak measurements of photon number of the
radiation field. Negative weak values of energy contradicts all
classical stochastic models, whereas negative weak values of photon
number contradict all classical stochastic models where the energy
is bounded from below by the zero-point energy. We examine coherent
states in particular, and find negative weak values with
probabilities of 16 \% for kinetic energy (or squared field
quadrature), 8 \% for harmonic oscillator energy and 50 \% for
photon number. These experiments are robust against detector
inefficiency and thermal noise.

\end{abstract}

 \keywords{}

\maketitle

\section{Introduction}\label{sec:intro}

Physics is an endeavor to construct a mathematical model of
nature. The basic mathematical model of classical physics is one
in which all dynamical variables have definite values
simultaneously. With the advent of statistical mechanics, a
probabilistic description was introduced in which each dynamical
variable would have some value with a certain probability. This is
the most general description that classical physics can provide. A
great variety of phenomena can be explained in terms of such a
model. For example, most optical phenomena can be described in
terms of a model of complex, stochastic amplitudes. It was not
until 1977 that this model was found to break down in an
experiment which demonstrated the phenomenon of antibunching
\cite{Kimble+DagenaisETAL-PhotAntiResoFluo:77}.

In this paper we examine a relatively new type of measurement
known as weak measurements
\cite{Aharonov+AlbertETAL-ResuMeasCompSpin:88}. Weak measurements
may be performed in exactly the same way as standard von Neumann
measurements, but with a weakened interaction
\cite{Johansen-WeakMeaswithArbi:04}. In a weak measurement, the
pointer will on average register the expectation value of the
observable that is measured. However, when the weak measurement is
conditioned on a second postselection measurement, the pointer
registers the real part of what is known as the ``weak value" of
the observable. These weak values have caused a lot of
controversy, in particular because they may exceed the eigenvalue
spectrum of the observable.

The main objective of this paper is to discuss the conditions
under which a classical description of weak measurements is
possible, taking into account practical experimental limitations
and possibilities. In particular, our purpose is to investigate in
further detail the failure of providing a classical description of
coherent states in such experiments, and to propose a feasible
experiment demonstrating nonclassical properties of coherent
states.

In Sec. \ref{sec:glauber} we give a brief review of the Glauber
classicality criterion, and discuss it's possible limitations.

In Sec. \ref{sec:weak} and appendices \ref{app:povm} and
\ref{app:weak} we consider a general class of weak measurements
where the object may be prepared in a mixed state, and where the
pointer may be prepared in an arbitrary mixed state of vanishing
current density. We consider detectors of arbitrary quantum
efficiency, and derive a generalized weak value.

In Sec. \ref{sec:classical} we demonstrate that the weak value can
be expressed as a conditional expectation over an infinite set of
different quasi-probability distributions. These distributions can
be regarded as generalizations of the complex Kirkwood
distribution \cite{Kirkwood-QuanStatAlmoClas:33} or the standard
ordered distribution \cite{Mehta-PhasFormDynaCano:64}. We denote
such distributions by $S$-distributions. The
Terletsky-Margenau-Hill distribution
\cite{Terletsky-claslimiquanmech:37,%
Margenau+Hill-CorrbetwMeasQuan:61}, or $T$-distribution, is the real
parts of the $S$-distribution. We find that strange weak values (for
which the real part exceeds the eigenvalue spectrum of the
observable) can only exist if the $T$-distribution takes negative
values. We demonstrate that a classical model of a weak measurement
exists whenever the representation of the observable coincides with
the classical representation of the observable and the
$T$-distribution is nonnegative. We demonstrate in particular that
negative weak values of energy contradict all classical stochastic
models, and that negative weak values of photon number contradict a
stochastic model where the energy is bounded from below by the
zero-point energy.

In Sec. \ref{sec:experimental} we propose two feasible practical
realizations of weak measurements of photon number and energy.

In Sec. \ref{sec:coherent} we consider in particular weak
measurements on coherent states. We demonstrate that coherent states
display negative weak values with probabilities of 16 \% for kinetic
energy (or squared field quadrature), 8 \% for harmonic oscillator
energy and 50 \% for photon number. We find that these effects are
robust against detector inefficiency and thermal noise.

\section{The Glauber classicality criterion}\label{sec:glauber}

Glauber and Sudarshan have demonstrated that any density operator
can be expanded diagonally in terms of coherent states
\cite{Glauber-CoheIncoStatRadi:63,Sudarshan-EquiSemiQuanMech:63}
\begin{equation}\label{eq:pdistr}
    \hat{\rho} = \int d^2 \alpha P(\alpha) \mid \alpha \rangle
    \langle \alpha \mid.
\end{equation}
The weight function $P(\alpha)$ is known as the $P$-distribution.
Furthermore, one may write the expectation value of any normal
ordered operator $\mathcal N \left [ \hat{O} (\hat{a}^\dag,
\hat{a}) \right ]$ as an integral of the form
\begin{equation}
    \langle \mathcal N \left [ \hat{O} (\hat{a}^\dag, \hat{a})
    \right ] \rangle = \int d^2 \alpha O (\alpha^*, \alpha) P(\alpha),
\end{equation}
where $a^*$ is the $c$-number representation of $\hat{a}^\dag$ and
$a$ is the $c$-number representation of $\hat{a}$. Therefore, if
the $P$-distribution has the properties of a valid probability
distribution, one may say that a classical representation exists
for any normal ordered operator product. This is known as the
``optical equivalence theorem"
\cite{Sudarshan-EquiSemiQuanMech:63}.

The optical equivalence theorem is the basis of the Glauber
classicality criterion, according to which all states for which
the $P$-distribution is a probability distribution are regarded as
essentially classical. If $P$ fails to be a probability
distribution, the state is considered as nonclassical
\cite{Glauber-CoheIncoStatRadi:63,%
Titulaer+Glauber-CorrFuncCoheFiel:65,%
Hillery-Claspurestatcohe:85,%
Mandel+Wolf-OptiCoheQuanOpti:95,%
Dodonov-Noncstatquanopti:02}. The Glauber criterion is widely
accepted as giving the most general distinction between quantum
and classical states. It is the basis of various measures of
``nonclassicality". For example, Hillery has defined a measure of
nonclassicality based on the distance in Hilbert space between the
object state and coherent states
\cite{Hillery-Noncdistquanopti:87}. Lee has defined a nonclassical
depth defined as the minimum average number of thermal photons
that must be added to render the $P$-distribution nonnegative
\cite{Lee-Measnoncnoncstat:91}. More recently, Vogel \emph{et.al}
have derived a hierarchy of observable conditions to test the
Glauber criterion
\cite{Vogel-NoncStat:00,Richter+Vogel-NoncQuanStat:02}.

The Glauber criterion must be considered a conjecture rather than a
proven theorem. It is based on some plausible arguments. Firstly,
due to the multitude of arguments in favor of coherent states as the
only classical-like pure states, it is natural to assume that also
classical mixtures of coherent states are classical. This is in fact
equivalent to Glauber's classicality criterion, as can be seen from
Eq. (\ref{eq:pdistr}). If the $P$-distribution is a probability
distribution, the density matrix can be expressed as a classical
mixture of coherent states. Secondly, a nonnegative $P$-distribution
ensures that the whole range of $s$-ordered Wigner distributions are
also nonnegative. Therefore, the $P$-distribution is most
``sensitive" to nonclassicality among all $s$-ordered distributions.

It is puzzling that a definition of nonclassicality depends on the
ability to represent expectations of \emph{normally ordered}
operator expressions as classical expectation values over a
probability distribution. This is sometimes attributed to the fact
that normal ordering of operators is closely associated to the
theory of photo-detection.

However, we may devise experimental procedures related to operator
orderings different from normal ordering of annihilation and
creation operators, that may display clear contradictions between
the classical and quantum descriptions for the same experiment.
This is actually the case of weak measurements, whose statistics
are related to $T$-distributions instead of to the more standard
$s$-ordered distributions. We show below that weak measurements
lead to strange outputs provided that the $T$-distribution takes
negative values. Moreover, we will show in this paper that a
classical stochastic model may fail to describe weak measurements
even when the $P$-distribution is a nonnegative probability
distribution.

\section{Weak measurements}\label{sec:weak}

A ``measurement" comes about when an auxiliary system interacts with
an object. By examining the properties of the auxiliary system after
the interaction, it may be possible to extract information about the
object. The auxiliary system is frequently called a ``measurement
apparatus" or a ``pointer system". The basic theory of quantum
measurement was examined by von Neumann in his seminal work on the
mathematical foundations of quantum mechanics
\cite{Neumann-MathFounQuanMech:55}. In this work, von Neumann
represented the interaction between the object and the pointer by an
interaction Hamiltonian of the form (throughout we will use units in
which $\hbar=1$)
\begin{equation}\label{eq:quantuminteraction}
    \hat{H}_\epsilon = \epsilon \delta(t) \; \hat{\nu} \otimes
    \hat{P}.
\end{equation}
A short explanation of the terms involved is in place. The
constant $\epsilon$ represents the interaction strength. The
interaction is of short duration, represented in idealized form by
the $\delta$-term. The hermitian observable $\hat{\nu}$ belongs to
the object Hilbert system $\mathcal{H}_s$, and is the observable
that we want to ``measure". The observable $\hat{P}$ is the
pointer momentum, and belongs to the pointer Hilbert space ${\cal
H}_a$. Although seemingly artificially constructed, this
interaction model has served as an archetype of the interaction
mechanism in quantum measurements. It has been found that the
conclusions that can be drawn from this model are generic to a
number of other interaction mechanisms (for a closer discussion of
the specific properties of this interaction Hamiltonian, see Ref.
\cite{Haake+Walls-Overamplmetequan:87}).

In a standard, projective measurement, the pointer position
$\hat{Q}$, with $[\hat{Q},\hat{P}]=i$, displays one of the
eigenvalues of the object observable $\hat{\nu}$ after the
measurement interaction. It was demonstrated by von Neumann that
in order to accomplish this, the state of the pointer prior to the
interaction should have a small position spread
\cite{Neumann-MathFounQuanMech:55}. The same effect can be
accomplished by allowing the interaction strength $\epsilon$ to be
sufficiently large \cite{Haake+Walls-Overamplmetequan:87}. For
this reason, this type of measurement is frequently called a
``strong measurement".

Until recently, it was thought that strong measurements are the
only useful type of measurements in quantum mechanics. However, in
1988 Aharonov \emph{et.al.} proposed a new type of measurement
that they called ``weak measurements"
\cite{Aharonov+AlbertETAL-ResuMeasCompSpin:88}. Such measurements
also employ the von Neumann interaction mechanism
(\ref{eq:quantuminteraction}), the difference being that the
pointer is assumed to be in an initial state of large position
uncertainty. More specifically, they assumed that the initial
state of the pointer was a Gaussian with large spread.

Recently, it was shown that weak measurements can be performed
also when the pointer is in an arbitrary mixed state, provided
that the interaction strength $\epsilon$ is sufficiently small and
the current density of the pointer vanishes
\cite{Johansen-WeakMeaswithArbi:04}. This description can be
generalized further by taking into account finite efficiency of
the detectors. A detector of finite efficiency can be represented
by Positive Operator Valued Measure (POVM) (see Appendix
\ref{app:povm}). In this case, the weak value of the observable
$\hat{\nu}$ conditioned on an imperfect postselection of the
observable $\hat{\phi}$ on the object is (see Appendix
\ref{app:weak})
\begin{equation}\label{eq:weakvalue}
    \nu_w(\phi) = {\mathrm{Tr} ( \hat{\Pi}_\phi \hat{\nu}
    \hat{\rho}_s ) \over \mathrm{Tr} (\hat{\Pi}_\phi
    \hat{\rho}_s)},
\end{equation}
where $\hat{\Pi}_\phi$ is a diagonal POVM representing the
imperfect postselection and $\hat{\rho}_s$ is the initial state of
the object (see Eq. (\ref{eq:objectpovm})).

\section{On classical models of weak
measurements}\label{sec:classical}

Under what circumstances is it possible to find a classical
representation of a weak measurement? Or put differently, under what
circumstances can the outcome of a weak measurement be modelled in
terms of a classical, stochastic model? By answering this question,
we will also understand under what circumstances weak measurements
demonstrate nonclassical properties of the quantum state under
consideration.

We begin by inserting the definition (\ref{eq:objectpovm}) for the
diagonal postselection POVM into Eq. (\ref{eq:weakvalue}). We then
have
\begin{equation}
    \nu_w (\phi) = {\mathrm{Tr} \left [ \int d\phi' \; \Pi_\phi(\phi')
    \mid \phi' \rangle \langle \phi' \mid \hat{\nu}
    \hat{\rho}_s \, \right ] \over \mathrm{Tr} \left [ \int d\phi' \;
    \Pi_\phi(\phi') \mid \phi' \rangle \langle \phi' \mid \hat{\rho}_s
    \, \right ]}.
\end{equation}
By performing the trace over any complete set of states we obtain
\begin{equation}\label{eq:weaktemp}
    \nu_w (\phi) = {\int d\phi' \; \Pi_\phi(\phi') \langle \phi' \mid
    \hat{\nu} \hat{\rho}_s \mid \phi' \rangle \over \int d\phi' \;
    \Pi_\phi(\phi') \langle \phi' \mid \hat{\rho}_s \mid \phi' \rangle}.
\end{equation}
We employ an arbitrary complete set of states $\mid \xi \rangle$,
\begin{equation}\label{eq:xicomplete}
    \int d\xi \mid \xi \rangle \langle \xi \mid = \mathrm{I}.
\end{equation}
We may then write Eq. (\ref{eq:weaktemp}) in the form
\begin{equation}\label{eq:gencond}
    \nu_w(\phi) = {\int d\xi \, d\phi' \, \Pi_\phi(\phi') \;
    S_{\hat{\nu}}(\phi',\xi) S(\phi',\xi)\over \int d\xi \,
    d\phi' \; \Pi_\phi(\phi') S(\phi',\xi)}.
\end{equation}
where
\begin{equation}
    S_{\hat{\nu}}(\phi,\xi) = {\langle \phi | \hat{\nu} | \xi
    \rangle \over \langle \phi | \xi \rangle},
\end{equation}
is a $c$-number representation of the observable $\hat{\nu}$. In
fact, it is the weak value for the observable preselected in the
state $\mid \xi \rangle$ and postselected in the state $\mid \phi
\rangle$. Also,
\begin{equation}
    S(\phi,\xi) = \langle \phi | \xi \rangle \langle \xi |
    \hat{\rho}_s | \phi \rangle
\end{equation}
is a generalization of the Kirkwood distribution for arbitrary
observables $\hat{\xi}$ and $\hat{\phi}$
\cite{Kirkwood-QuanStatAlmoClas:33}. The Kirkwood distribution is
also known as the anti-standard ordered distribution
\cite{Mehta-PhasFormDynaCano:64}. $S$ is also a generalization of
the standard ordered distribution, which is the complex conjugate
of the Kirkwood distribution \cite{Mehta-PhasFormDynaCano:64}. In
this paper, we will simply denote it by the $S$-distribution, and
we will refer to $S_{\hat{\nu}}$ as the $S$-representation of the
observable $\hat{\nu}$.

The $S$-distributions are in general complex, and as such are
quasi-probability distributions. Nevertheless, they possess some
of the properties of classical joint distributions. For example,
assuming that both eigenstates $\mid \phi \rangle$ and $\mid \xi
\rangle$ constitute complete sets, it is straightforward to show
that they yield correct marginal distributions when integrated
over either variable,
\begin{eqnarray}
    \langle \phi \mid \hat{\rho}_s \mid \phi \rangle &=& \int d\xi
    \; S(\phi,\xi),\\
    \langle \xi \mid \hat{\rho}_s \mid \xi \rangle &=& \int d\phi
    \; S(\phi,\xi).
\end{eqnarray}
It is straightforward to show that also the complex conjugate
distribution, $S^*(\phi,\xi)$, fulfills such marginality
conditions.

The $S$-distribution can be expressed in the form
\begin{equation}\label{eq:weakdensity}
    S(\phi,\xi) = \mid \langle \phi | \xi \rangle \mid^2 {\langle \xi |
    \hat{\rho}_s | \phi \rangle \over \langle \xi | \phi \rangle}.
\end{equation}
In this form, the $S$-distribution is a product of a nonnegative
probability distribution and the weak value of the density operator.
If $\mid \xi \rangle$ or $\mid \phi \rangle$ are eigenstates of the
density operator, $S$ will be real and nonnegative.

Classically, the weak value of an observable is the conditional
expectation of that observable
\cite{Johansen-WeakMeaswithArbi:04}. The expression
(\ref{eq:gencond}) demonstrates that there exists an infinite set
of representations under which the weak value can be expressed as
a conditional expectation of a $c$-number variable $S_{\hat{\nu}}$
over an $S$-distribution. For each choice of basis $\mid \xi
\rangle$, a different $S$-distribution is obtained. However, some
of these representations bear little resemblance to any classical
model. For example, in some models the $S$-representation of the
hermitian observable $\hat{\nu}$ is complex. As a basic
requirement on a \emph{classical} model, we shall in the following
restrict the attention to the subset of representations for which
hermitian observables have real representations,
\begin{equation}\label{eq:realc}
    \mathrm{Im} S_{\hat{\nu}}(\phi,\xi) = 0.
\end{equation}
We also introduce the $T$-representation of $\hat{\nu}$ and the
$T$-distribution \cite{Terletsky-claslimiquanmech:37,%
Margenau+Hill-CorrbetwMeasQuan:61},
\begin{eqnarray}
  T_{\hat{\nu}}(\phi,\xi) &=& \mathrm{Re} S_{\hat{\nu}}(\phi,\xi), \\
  \label{eq:realpart}
  T(\phi,\xi) &=& \mathrm{Re} S (\phi,\xi).
\end{eqnarray}
It is straightforward to show that also these provide correct
marginal distributions,
\begin{eqnarray}
    \langle \phi \mid \hat{\rho}_s \mid \phi \rangle &=& \int d\xi
    \; T(\phi,\xi),\\
    \langle \xi \mid \hat{\rho}_s \mid \xi \rangle &=& \int d\phi
    \; T(\phi,\xi).
\end{eqnarray}
Because of the classicality condition (\ref{eq:realc}), the $S$-
and $T$-representations of the observable are the same,
\begin{equation}\label{eq:tmhvar}
    T_{\hat{\nu}}(\phi,\xi) = {\langle \phi | \hat{\nu} | \xi
    \rangle \over \langle \phi | \xi \rangle}.
\end{equation}
The distribution of the postselection observable $\hat{\phi}$,
taking into account the finite detector efficiency represented by
$\Pi_\phi$, can be found both from the complex $S$-distribution
and the real $T$-distribution through the integrals
\begin{eqnarray}
\mathrm{Tr} \left ( \Pi_\phi  \hat{\rho}_s \right ) &=& \int d\xi
\, d\phi' \; \Pi_\phi(\phi') S(\phi',\xi) \nonumber \\
&=& \int d\xi \, d\phi' \; \Pi_\phi(\phi') T(\phi',\xi).
\end{eqnarray}
In a weak measurement, the real part of the weak value $\nu_w$ is
registered by the pointer. Under the assumption (\ref{eq:realc}),
we may write
\begin{equation}\label{eq:realweak}
    \mathrm{Re} (\nu_w ) = {\int d\xi \, d\phi' \, \Pi_\phi(\phi') \;
    T_{\hat{\nu}}(\phi',\xi) T(\phi',\xi)\over \int d\xi \,
    d\phi' \; \Pi_\phi(\phi') T(\phi',\xi)}.
\end{equation}
This expression, which reflects the expectation of the pointer
displacement in a weak measurement, bears formal resemblance to a
classical conditional expectation. The difference is that the
$T$-distribution may take negative values, and that the
$T$-representation of the observable may differ strongly from the
representation of the observable in classical theory. A classical
representation of the pointer displacement can be said to exist if
$T_{\hat{\nu}}$ equals the classical representation of the
observable and if also the $T$-distribution is nonnegative.

What determines the sign of the $T$-distribution? From Eqs.
(\ref{eq:weakdensity}) and (\ref{eq:realpart}) it follows that
\begin{equation}
    T(\phi,\xi) = \mid \langle \phi | \xi \rangle \mid^2 \mathrm{Re}
    \left ( {\langle \xi | \hat{\rho}_s | \phi \rangle \over \langle
    \xi | \phi \rangle} \right ).
\end{equation}
Thus, the sign of the $T$-distribution equals the sign of the real
part of the weak value of the density operator. Therefore, a
requirement for nonclassicality is that the real part of the weak
value of the density operator should be negative.

In the following, we consider two representations that may provide
classical-like models of weak measurements. The first
representation, which we shall call the \emph{eigenvalue
representation}, is found when the basis $\mid \xi \rangle$ is
chosen as eigenstates of the observable $\hat{\nu}$. In this
representation the $S$-and $T$-representations of the observable
$\hat{\nu}$ are the eigenvalues $\nu$,
\begin{equation}
    S_{\hat{\nu}}(\phi,\nu) = T_{\hat{\nu}}(\phi,\nu) = \nu,
\end{equation}
By using this representation, we have
\begin{eqnarray}
    \nu_w (\phi) &=& \int d\nu \, \nu \, S_\eta(\nu \mid \phi),\\
    \label{eq:conditional}
    \mathrm{Re} \nu_w (\phi) &=& \int d\nu \, \nu \, T_\eta(\nu \mid
    \phi),
\end{eqnarray}
where
\begin{eqnarray}\label{eq:cont}
    S_\eta(\nu \mid \phi) &=& {S_\eta(\nu, \phi) \over \int d\nu
    S_\eta(\nu, \phi)},\\
    T_\eta(\nu \mid \phi) &=& {T_\eta(\nu, \phi) \over \int d\nu
    T_\eta(\nu, \phi)}
\end{eqnarray}
are ``effective", conditional distributions, and where
\begin{eqnarray}\label{eq:convolution}
    S_\eta(\phi,\nu) &=& \int \, d\phi' \; \Pi_\phi(\phi')
    S(\phi',\nu),\\
    T_\eta(\phi,\nu) &=& \int \, d\phi' \; \Pi_\phi(\phi') T(\phi',\nu)
\end{eqnarray}
are ``effective" $S$- and $T$-distributions for $\nu$ and $\phi$.

It follows straightforwardly from Eq. (\ref{eq:conditional}) that
``strange weak values" where the real part of $\nu_w$ exceeds the
eigenvalue spectrum is only possible for quantum states for which
the $T$-distribution in the eigenvalue representation takes
negative values. But it is of course well known that classical
models may allow observables to exceed the eigenvalue spectrum of
the observable. This means also that ``strange" weak values may
sometimes by supported by a classical model. In particular, this
may be the case for observables with a discrete spectrum.

The other $c$-number representation that we shall consider is the
\emph{phase space representation}. In this case, the basis $|\,\xi
\, \rangle$ should be chosen as eigenstates of the observable
canonically conjugate to the postselection observable
$\hat{\phi}$. This may provide the possibility of a comparison
with a classical phase space description of weak measurements.

We give a couple of examples illustrating the use of these two
representations. First, we consider a weak measurement of a field
squared quadrature (or kinetic energy for material particles)
$\hat{\nu} = \hat{p}^2$ postselected on the canonically conjugate
field quadrature (or position for material particles) $\hat{\phi} =
\hat{q}$
\cite{Aharonov+PopescuETAL-MeasErroNegaKine:93,%
Johansen-Noncpropcohestat:04}.
In this case, the eigenvalue representation and the phase space
representation are one and the same, since $\hat{q}$ and $\hat{p}$
are canonically conjugate variables. From Eq. (\ref{eq:tmhvar})
follows that both the eigenvalue representation and the phase space
representation of $\hat{p}^2$ postselected on $\hat{q}$ is
\begin{equation}\label{eq:kinetic}
    T_{\hat{p}^2}(q,p) = p^2.
\end{equation}
Obviously, $T_{\hat{p}^2}(q,p) \ge 0$. From Eq.
(\ref{eq:conditional}) and the classicality assumption $T(q,p) \ge
0$ follows the inequality
\begin{equation}
    \mathrm{Re} \, (p^2_w)(q)  \ge 0.
\end{equation}
In this case, a ``strange" negative weak value implies failure of
the classical model where $p^2$ takes the positive continuum. It can
be noted that a negative weak value of kinetic energy $\hat{p}^2$
contradicts \emph{all} stochastic $c$-number models where kinetic
energy takes only nonnegative values, even models where the
$T$-representation of kinetic energy may differ from the expression
(\ref{eq:kinetic}).

Consider next a weak measurement of the energy of a harmonic
oscillator (we use units so that $\omega = 1$)
\begin{equation}
    \hat{H} = {1 \over 2} \left ( \hat{p}^2 + \hat{q}^2 \right ) =
    \hat{n} + {1 \over 2},
\end{equation}
assuming postselection on one of the quadratures (or position for a
material particle). In the eigenvalue representation
\begin{equation}\label{eq:discrete}
    T_{\hat{H}}(q,n) = n + {1 \over 2}.
\end{equation}
Obviously, $T_{\hat{H}}(q,n) \ge {1 \over 2}$. In the eigenvalue
representation, the $T$-representation of the Hamiltonian is bounded
from below by the zero-point energy. From Eq. (\ref{eq:conditional})
and by imposing the classicality criterion $T(q,n) \ge 0$, we may
derive the inequality
\begin{equation}\label{eq:strange}
 \mathrm{Re} (H_w)(q) \ge {1 \over 2},
\end{equation}
Violation of this inequality implies ``strange" weak values. It can
only take place if the $T$-distribution $T(q,n)$ takes negative
values. We can split the strange values into two categories. (i)
Outputs of the form ${1 \over 2} \geq \mathrm{Re} (H_w)(q) \ge 0$
contradict exclusively those classical stochastic models for which
the excitation number takes positive values $n \geq 0$, or,
equivalently, for which the energy is larger than or equal to $1/2$.
These outputs do not contradict classical models where the energy
takes positive values starting from zero. (ii) On the other hand,
outputs of the form $0 >\mathrm{Re} (H_w)(q)$ contradict all the
classical stochastic models where the energy takes positive values.

In the phase-space representation, a weak measurement of the energy
of a harmonic oscillator postselected on one of the quadratures
calls for the $T$-representation \begin{equation}
    T_{\hat{H}}(q,p) = {1 \over 2} \left ( p^2 + q^2 \right ). \end{equation} This representation of harmonic oscillator energy coincides with the classical representation, and therefore permits an investigation of the limits of classical models. Obviously we have $T_{\hat{H}}(q,p) \ge 0$. By assuming that also the state representation is classical, $T(q,p) \ge 0$, one may derive from Eq. (\ref{eq:realweak}) the inequality \begin{equation}\label{eq:continuousmodel}
    \mathrm{Re} (H_w)(q) \ge 0.
\end{equation}
This inequality should be compared with inequality
(\ref{eq:strange}). We may say that the phase-space representation
highlights just the second category (ii) above: Violation of
inequality (\ref{eq:continuousmodel}) rules out all the classical
stochastic models where the energy takes positive values. Note that
the probability of infringing inequality (\ref{eq:continuousmodel})
and the probability of the second category (ii) above are equal,
since the probability of $\mathrm{Re} (H_w)(q) < 0$ does not depend
on the representation. On the other hand, the probability that
$\mathrm{Re} (H_w)(q) < 0$ is less than the probability of
$\mathrm{Re} (H_w)(q) < 1/2$ in agreement with the fact that the
first one excludes a larger class of classical models.

We have not mentioned the correspondence principle here. Our purpose
is to investigate under what conditions quantum mechanics can be
reproduced by a classical stochastic theory. This is not related
directly to the classical limit of quantum mechanics.

\section{Weak measurement of photon number}
\label{sec:experimental}

The realization of weak measurements requires the coupling of the
system to be observed with auxiliary degrees of freedom. The
output of the weak measurement is inferred from measurements
carried out on the auxiliary system and on the object system
itself. In this work we will consider the weak measurement of two
observables with nonnegative spectra. The boundedness of the
spectra is mandatory in order to reveal the appearance of strange
values. These observables are $\hat{\nu}=\hat{p}^2$ and the number
operator $\hat{\nu}=\hat{a}^\dagger \hat{a}$, where
$\hat{a}=(\hat{q}+i\hat{p}) /\sqrt{2}$ and $\hat{q}$, $\hat{p}$,
with $[\hat{q}, \hat{p}]=i$ are the standard position and linear
momentum or field quadratures. For both examples of $\hat{\nu}$ we
will consider the same postselection strategy given by the
measurement of the operator $\hat{\phi}= \hat{q}$.

In this section we propose two simple and feasible schemes for the
weak measurement of $\hat{a}^\dagger \hat{a}$ conditioned on the
measurement of $\hat{q}$ in the field of quantum optics, where
$\hat{a}^\dagger \hat{a}$ represents the number of photons and
$\hat{q}$ is a field quadrature. The two possibilities involve
different realizations of the auxiliary system. These are another
field mode and two-level atoms.

\subsection{Coupling to a field mode}

Let us assume that the auxiliary system is another field mode with
complex amplitude operator $\hat{b}$. A suitable coupling between the
system $\hat{a}$ and the auxiliary variables allowing a weak measurement
of the number operator $\hat{a}^\dagger \hat{a}$ is of the form
\begin{equation}
H_\epsilon = \epsilon \hat{a}^\dagger \hat{a} \hat{b}^\dagger \hat{b}.
\end{equation}

\begin{figure}
\includegraphics[width=6cm]{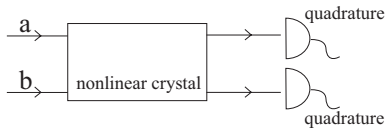}
\caption{Scheme for a weak measurement via cross-Kerr coupling in
a nonlinear crystal.}
\end{figure}

This coupling can be achieved in practice by propagation of both
field modes in crystals with nonlinear optical properties
(cross-Kerr interaction) so that $\epsilon$ is proportional to the
nonlinear susceptibility of the medium and the length of the
crystal. This coupling causes a phase shift of the mode $\hat{b}$
proportional to the photon number in mode $\hat{a}$, that can be
then detected simply by measuring a quadrature of the mode
$\hat{b}$ (homodyne detection) \cite{Walls+Milburn-QuanOpti:94,%
Scully+Zubairy-QuanOpti:97}. This example has the advantage that
nonlinear effects are usually very weak, so that the requirement
$\epsilon \rightarrow 0$ for a weak measurement is naturally
satisfied. In Fig. 1 we outline the scheme of the weak measurement
of photon number via cross-Kerr coupling conditioned on a
quadrature measurement.

\subsection{Coupling to a two-level atom}

The weak measurement of the photon number can also be carried out
by coupling the field mode to a two-level atom with internal
energy levels $|\pm \rangle$. If the frequency of the field and
the resonant frequency of the atom are detuned enough the
atom-field interaction Hamiltonian becomes
\cite{Walls+Milburn-QuanOpti:94,Scully+Zubairy-QuanOpti:97}
\begin{equation}
H_\epsilon = \epsilon \hat{a}^\dagger \hat{a} \sigma_z ,
\end{equation}
where $\sigma_z = | + \rangle \langle + | - | - \rangle \langle -
|$. In this case we have that $\epsilon$ is inversely proportional
to the detuning, so the condition $\epsilon \rightarrow 0$ can be
easily achieved.

The atom-field interaction causes a phase shift of the
coefficients of an atomic superposition of the states $|\pm
\rangle$. The shift is proportional to the photon number and can
be detected by measuring, for example, the observable
\begin{equation}
\sigma_x = | + \rangle \langle - | + | + \rangle \langle - | .
\end{equation}
The measurement of $\sigma_x$ can be carried out by determining
the population of the levels $|\pm \rangle$ after applying to the
atom a resonant pulse transforming the eigenstates $| \pm
\rangle_x$ of $\sigma_x$ into $| \pm \rangle$.

The condition of null current density (\ref{eq:zerocurrent})
becomes in this case
\begin{equation}
{}_x \langle \pm | \left ( \sigma_z \rho_a + \rho_a \sigma_z
\right ) | \pm \rangle_x = 0 ,
\end{equation}
which is verified always provided that
\begin{equation}
\rho_a = \frac {1}{2} \left ( 1 + s_x \sigma_x + s_y \sigma_y
\right ) ,
\end{equation}
where $s_x$, $s_y$ are real constants such that $s_x^2 + s_y^2
\leq 1$ and
\begin{equation}
\sigma_y = i \left ( | + \rangle \langle - | - | + \rangle \langle
- | \right ) .
\end{equation}

Besides quantum optics, this scheme can be also implemented in the
context of trapped ions, where $\hat{a}$ would represent the complex
amplitude of the one-dimensional harmonic motion of the center of
mass of the trapped ion, and $|\pm \rangle$ are two internal
levels of the same ion.

\section{Coherent states}\label{sec:coherent}

In this section, we turn to the study of weak measurements on
coherent states, taking into consideration the effect of thermal
noise and finite detector efficiency. It is of particular interest
to study possible nonclassical properties of coherent states,
since coherent states are the only pure states that satisfy the
Glauber criterion of possessing a nonnegative $P$-distribution.

We shall consider the thermalized coherent state (also known as
the displaced thermal state)
\begin{equation}
    \hat{\rho} = \hat{D}(\alpha) \hat{\rho_{th}}
    \hat{D}^\dag(\alpha).
\end{equation}
Here $\hat{\rho_{th}}$ is the density operator for the thermal
state and $\hat{D}(\alpha)$ is the displacement operator. This
state has a nonnegative $P$-distribution
\cite{Lachs-TheoAspeMixtTher:65}
\begin{equation}
    P(\gamma) = {1 \over \pi n_{th}} e^{-\mid \gamma - \alpha
    \mid^2/n_{th}},
\end{equation}
where $\alpha$ is the coherent amplitude when $n_{th}$ vanishes
and $n_{th}$ is the expected thermal photon number when $\alpha$
vanishes. Since the $P$-distribution is nonnegative, this state is
essentially classical according to the Glauber criterion.

In this section, we consider weak measurements with postselection
on position. The phase space for this experiment consists of
position and momentum. In the phase space representation, the
$S$-distribution for this state is
\begin{equation}
    S(q,p) = S_{th}(q-\alpha_r,p-\alpha_i) ,
\end{equation}
where $\alpha = (\alpha_r + i \alpha_i)/\sqrt{2}$,
\begin{equation}\label{eq:thermal}
    S_{th}(q,p) ={\exp \left [- {2 \sigma_{th}^2 (p^2+q^2) -
    2 i p q \over {1  + 4 \sigma_{th}^4}} \right ] \over \pi
    \sqrt{1 + 4 \sigma_{th}^4}},
\end{equation}
is the $S$-distribution of a thermal state
\cite{Johansen-NoncTherRadi:04} and
\begin{equation}
    \sigma_{th}^2 = n_{th} + {1 \over 2}
\end{equation}
is the variance of each quadrature for the thermal distribution.
Clearly, the $T$-distribution for this state takes negative
values. It is worth emphasizing that the lack of positivity
persists for every $n_{th}$, in sharp contrast to the case of
s-ordered distributions for which there is always a value of
$n_{th}$ that renders the distribution positive.

In the following, we assume an imperfect measurement of the
operator $\hat{q}$ represented by a Gaussian postselection POVM
(see also Eq. (\ref{eq:objectpovm}))
\begin{equation}\label{gaussianpovm}
    \Pi_q (q') = {1 \over \sqrt{2 \pi} \sigma_\eta}
    {e^{-(q-q')^2/(2
    \sigma_\eta^2)}},
\end{equation}
where the width $\sigma_\eta$ is determined by detector
efficiencies. For example, for a homodyne detector, $\sigma_\eta^2
= (1-\eta)/(2\eta)$, where $\eta$ is the quantum efficiency of a
single detector. We may then derive an effective marginal
distribution for the postselection observable
\begin{equation}
    \rho_\eta(q) = \mathrm{Tr} \left [ \hat{\Pi}_q \hat{\rho_s}
    \right ] =
    {e^{-{(q-\alpha_r)^2 \over  2 (\sigma_{th}^2 +
    \sigma_\eta^2)}} \over \sqrt{2\pi(\sigma_{th}^2 +
    \sigma_\eta^2)}}.
\end{equation}

\subsection{Negative weak value of kinetic energy}

In this subsection, we consider weak measurements of the observable
$\hat{p}^2$, which is essentially the kinetic energy
\cite{Aharonov+PopescuETAL-MeasErroNegaKine:93}. The weak value of
$\hat{p}^2$ conditioned on the measurement of $\hat{q}$ was found to
take negative values for coherent states in Ref.
\cite{Johansen-Noncpropcohestat:04}. Here, this treatment will be
generalized to include the effects of thermal noise and imperfect
detectors.

\begin{figure}
\psfrag{nth}{$n_{th}$} \psfrag{h}{$\eta$} \psfrag{P}{$P$}
\includegraphics[width=6cm]{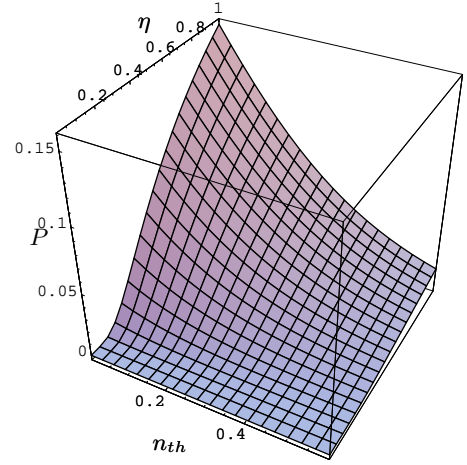}
\caption{The probability for observing a negative weak value of
$\hat{p}^2$ as a function of the detector efficiency $\eta$ and the
average thermal occupation number $n_{th}$. It is assumed that
$\alpha_i=0$, whereas $\alpha_r$ may take any value. This effect
therefore is present even for macroscopic excitations $\alpha_r$.}
\label{fig:p2wetanth}
\end{figure}

Since we are considering weak measurements of $\hat{p}^2$, the
phase space representation and the eigenvalue representation of
the $S$-distribution coincide. It is then useful to define an
effective $S$-distribution by using Eq. (\ref{eq:convolution}) and
(\ref{eq:thermal}). The result is
\begin{equation}
    S_\eta(q,p) = S_\eta^{th}(q-\alpha_r,p-\alpha_i),
\end{equation}
with
\begin{equation}\label{eq:thermaleffective}
    S_\eta^{th}(q,p) ={\exp \left [- {2 \sigma_{th}^2 (p^2+q^2)
    + 2 p^2 \sigma_\eta^2 - 2 i p q \over {1  + 4 \sigma_{th}^4} +
    4 \sigma_{th}^2 \sigma_\eta^2} \right ] \over \pi \sqrt{1 +
    4 \sigma_{th}^4 + 4 \sigma_{th}^2 \sigma_\eta^2}}
\end{equation}
being the effective $S$-distribution. Using Eq.
(\ref{eq:conditional}) and (\ref{eq:thermaleffective}) we find
that
\begin{equation}
    \mathrm{Re} \left [(p^2)_w \right ] = {1 + 4 (\alpha_i^2+\sigma_{th}^2)
    (\sigma_{th}^2 + \sigma_\eta^2) \over
    4 (\sigma_{th}^2 + \sigma_\eta^2)
    } - {(q-\alpha_r)^2 \over
    4 (\sigma_{th}^2 + \sigma_\eta^2)^2}.
\end{equation}
The roots of this polynomial are
\begin{equation}
    q_\pm = \alpha_r \pm \sqrt{(\sigma_{th}^2 + \sigma_\eta^2)[1
    + 4 (\alpha_i^2+\sigma_{th}^2)(\sigma_{th}^2+\sigma_\eta^2)]}.
\end{equation}
The probability of postselecting a position $q$ which on
average gives a negative weak value of $\hat{p}^2$ then is
\begin{equation}
    P \left [ \mathrm{Re}(p^2)_w<0 \right ] = 1-
    \int_{q_-}^{q_+} dq \, \rho_\eta(q).
\end{equation}
The result is
\begin{equation}
    P \left [ \mathrm{Re}(p^2)_w<0 \right ] = \mathrm{erfc}
    \sqrt{{1 \over 2} + 2(\alpha_i^2+\sigma_{th}^2)(\sigma_{th}^2
    +\sigma_\eta^2)}.
\end{equation}
The complementary error function $\mathrm{erfc}(x)$ is monotonically
decreasing. Therefore, this probability is maximized when
$\alpha_i^2$, $\sigma_{th}^2$ and $\sigma_\eta^2$ are as small as
possible. The state can be chosen so that $\alpha_i=0$. For a
perfect detector $\sigma_\eta=0$. The quadrature variance
$\sigma_{th}^2$ is bounded below by $1/2$. Therefore, the maximum
probability for a negative weak value of $\hat{p}^2$ is
$\mathrm{erfc} \, 1 \approx 0.16$
\cite{Johansen-Noncpropcohestat:04}. It is particularly interesting
to note that the effect is independent of the real part of the
amplitude, $\alpha_r$. A negative weak value therefore can be
observed for a macroscopic occupation of the mode.

This probability has been plotted as a function of detector
efficiency $\eta$ and thermal occupation number $n_{th}$ in Fig.
\ref{fig:p2wetanth}. As noted in section \ref{sec:classical}, the
negativity of $\mathrm{Re}(p^2_w)$ contradicts all classical
stochastic models. A weak value of kinetic energy might be observed
for material particles at low temperatures. However, a realizable
quantum optical experiment is not known to the authors. To
investigate a feasible quantum optical experiment, we turn to the
weak measurement of photon number and energy.

\subsection{Negative weak value of energy}

In this subsection we study weak measurements of energy
conditioned on the postselection of a quadrature observable. As
demonstrated in Sec. \ref{sec:classical}, a negative weak value of
energy for a harmonic oscillator contradicts classical stochastic
models.

\begin{figure}
\psfrag{ar}{$\alpha_r$} \psfrag{ai}{$\alpha_i$} \psfrag{P}{$P$}
\includegraphics[width=7cm]{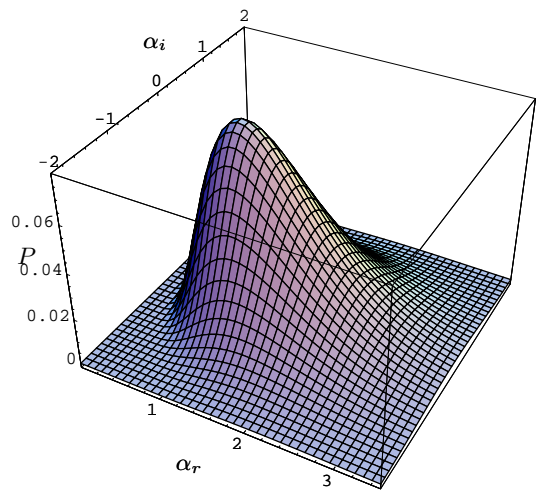}
\caption{The probability for observing a negative weak value of
energy as a function of the coherent state amplitudes $\alpha_r$ and
$\alpha_i$. We assume perfect detectors $\sigma_\eta=0$ and no
thermal noise $n_{th}=0$. The probability has a maximum at
$\alpha_r=1, \alpha_i = 0$, at which the probability is $(1/2) \,
\mathrm{erfc} \, 1 \approx 0.079$.} \label{fig:hwideal}
\end{figure}

By combining Eqs. (\ref{eq:realweak}), (\ref{eq:thermal}) and
(\ref{gaussianpovm}) we find that
\begin{equation}
    \mathrm{Re} \left [ H_w(q) \right ] = a q^2 + b q + c
\end{equation}
where
\begin{eqnarray}
  a &=& {4 \sigma_{th}^4-1 \over 8
  (\sigma_{th}^2+\sigma_\eta^2)^2},\\
  b &=& {\alpha_r (4 \sigma_{th}^2 \sigma_\eta^2 +1) \over 4
  (\sigma_{th}^2+\sigma_\eta^2)^2},\\
  c &=& {\sigma_{th}^2 \over 2} + {\alpha_i^2 \over 2} +
    {1 + 4  \sigma_{th}^2 \sigma_\eta^2 \over 8
    (\sigma_{th}^2+\sigma_\eta^2)} +
    {\alpha_r^2 (4 \sigma_\eta^4-1) \over 8
    (\sigma_{th}^2+\sigma_\eta^2)^2}.
\end{eqnarray}
Two real roots $q_\pm$ exist provided that $b^2 \ge 4 ac$ (only one
if  $a=0$). This establishes a necessary condition to be fulfilled
by $\sigma_{th}$, $\sigma_\eta$, $\alpha_r$, and $\alpha_i$ for the
existence of negative values for $\mathrm{Re} \left [ H_w(q) \right
]$. If this is satisfied, the probability of observing a negative
$\mathrm{Re} \left ( H_w \right )$ is
\begin{equation}
    P \left [ \mathrm{Re}(H_w)<0 \right ] = \int_{Q_-}^{Q_+} dq
    \, \rho_\eta(q).
\end{equation}
For ideal detectors $\sigma_\eta=0$ and vanishing thermal noise
$n_{th}=0$ this probability can be written as
\begin{equation}
    P \left [ \mathrm{Re}(H_w)<0 \right ] = {1 \over 2} \;
    \mathrm{erfc} \left ( {1 + \alpha_r^2+\alpha_i^2 \over 2 \mid
    \alpha_r \mid} \right ).
\end{equation}
This probability has been plotted in Fig. \ref{fig:hwideal}. It
reaches a maximum at $\alpha_r=1$ and $\alpha_i=0$, at which the
probability is $(1/2) \, \mathrm{erfc} \, 1 \approx 0.08$. This is
half of the probability for observing a negative weak value of
kinetic energy.

\begin{figure}
\psfrag{ar}{$\alpha_r$} \psfrag{ai}{$\alpha_i$} \psfrag{P}{$P$}
\includegraphics[width=7cm]{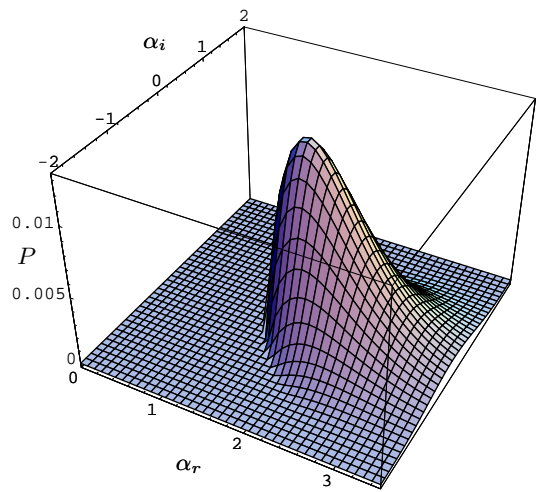}
\caption{The probability for observing a negative weak value of
energy as a function of the coherent state amplitudes $\alpha_r$ and
$\alpha_i$. We assume detectors with quantum efficiency $\eta=0.7$
and thermal occupation number $n_{th}=0.3$. There is a minimum
$\alpha_r$ required in order to see nonclassical behavior.}
\label{fig:hwtherm}
\end{figure}

\begin{figure}
\psfrag{nth}{$n_{th}$} \psfrag{h}{$\eta$} \psfrag{P}{$P$}
\includegraphics[width=6cm]{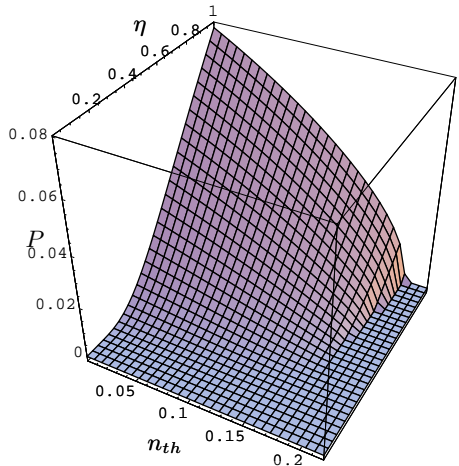}
\caption{The probability for observing a negative weak value of
energy as a function of the detector efficiency $\eta$ and the
average thermal occupation number $n_{th}$. The coherent excitation
is assumed to be $\alpha_r=1$, $\alpha_i=0$.} \label{fig:hwetanth}
\end{figure}

A more complex expression for the probability can be obtained for
arbitrary detector efficiency and thermal excitation. This has been
plotted in Fig. \ref{fig:hwtherm}. We see that $\alpha_r$ now must
reach a minimum value in order to see nonclassical negative weak
values. In Fig. \ref{fig:hwetanth}, the probability is plotted as a
function of detector efficiency $\eta$ and average thermal
occupation number $n_{th}$.

Negativity of weak values of kinetic energy and of total energy both
contradict a classical, stochastic model of light. Negative weak
values of kinetic energy persists also for macroscopic coherent
amplitudes. This does not occur for the weak value of energy. On the
other hand, a weak measurement of energy has a feasible practical
measuring schemes as outlined in Sec. \ref{sec:experimental}.

\subsection{Negative weak value of photon number}

It was demonstrated in Sec. \ref{sec:classical} that a negative weak
value of photon number contradicts a classical stochastic model
where the energy is bounded from below by the zero-point energy. We
study this further here.

\begin{figure} \psfrag{ar}{$\alpha_r$} \psfrag{ai}{$\alpha_i$}
\psfrag{P}{$P$}
\includegraphics[width=7cm]{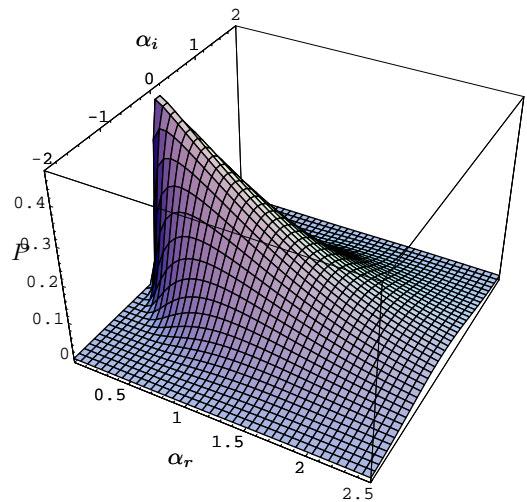}
\caption{The probability for observing a negative weak value of
the photon number $\hat{n}$ as a function of the coherent state
amplitudes $\alpha_r$ and $\alpha_i$. We assume perfect detectors
$\sigma_\eta=0$ and no thermal noise $n_{th}=0$. The probability
increases with decreasing $\mid \alpha \mid$, but has a
singularity at $\mid \alpha \mid = 0$.} \label{fig:nwideal}
\end{figure}

The weak value of photon number is simply
\begin{equation}
    n_w(q) = H_w(q) - {1 \over 2}.
\end{equation}
Two real roots $q_\pm$ of $n_w$ exist provided that $b^2 \ge 4
a(c-1/2)$. In the case that this condition is satisfied the
probability of observing a negative $\mathrm{Re} \left [n_w(q)
\right ]$ is
\begin{equation}
    P \left [ \mathrm{Re}(n_w)<0 \right ] = \int_{q_-}^{q_+} dq
    \, \rho_\eta(q).
\end{equation}
For ideal detectors and vanishing thermal noise this probability
can be written as
\begin{equation}\label{eq:nwideal}
    P \left [ \mathrm{Re}(n_w)<0 \right ] = {1 \over 2} \;
    \mathrm{erfc} \left ( {\alpha_r^2+\alpha_i^2 \over 2 \mid
    \alpha_r \mid} \right ).
\end{equation}
This function has been plotted in Fig. \ref{fig:nwideal}. The
probability is always maximized by letting $\alpha_i \rightarrow
0$. It approaches a maximum of 0.5 for vanishing $\alpha_r$.
However, it has a singularity in $\alpha_r=0$, and actually
vanishes in this point. Thus, there is zero probability of
observing a negative weak value of $\hat{n}$ for the
\textit{vacuum} state. There must be a finite \textit{small}
coherent amplitude to see this.

\begin{figure}
\psfrag{ar}{$\alpha_r$} \psfrag{ai}{$\alpha_i$} \psfrag{P}{$P$}
\includegraphics[width=7cm]{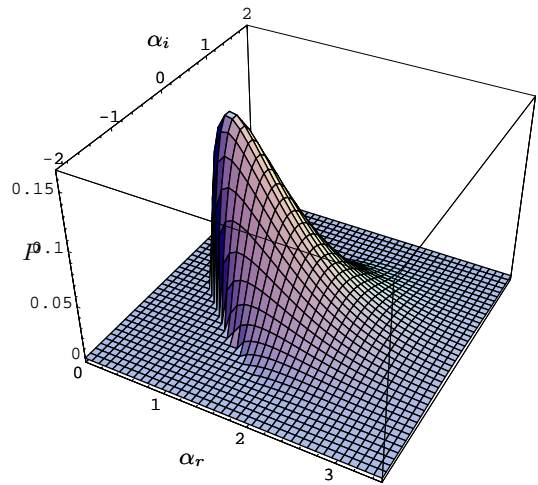}
\caption{The probability for observing a negative weak value of
the photon number $\hat{n}$ as a function of the coherent state
amplitudes $\alpha_r$ and $\alpha_i$. We assume detectors with
quantum efficiency $\eta=0.7$ and thermal occupation number
$n_{th}=0.3$. We see that there is a minimum $\alpha_r$ required
in order to see nonclassical behavior.} \label{fig:nwtherm}
\end{figure}

A more complex expression is obtained for finite detector
efficiency and finite thermal noise. The probability has been
plotted in this case in Fig. \ref{fig:nwtherm}. We see that
$\alpha_r$ now must reach a minimum value in order to see
nonclassical negative weak values.

\begin{figure}
\psfrag{nth}{$n_{th}$} \psfrag{h}{$\eta$} \psfrag{P}{$P$}
\includegraphics[width=6cm]{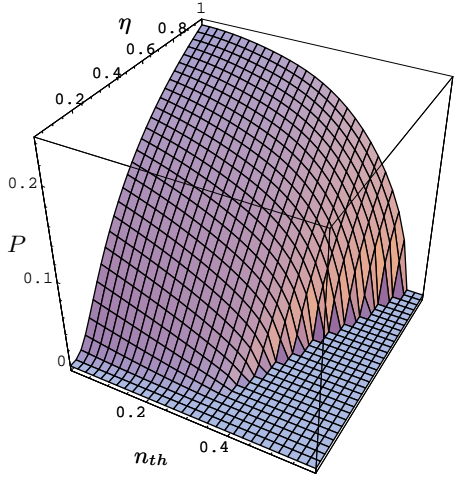}
\caption{The probability for observing a negative weak value of
the photon number $\hat{n}$ as a function of the detector
efficiency $\eta$ and the average thermal occupation number
$n_{th}$. The coherent excitation is assumed to be $\alpha_r=0.1$,
$\alpha_i=0$. A minimum detector efficiency (and a maximum thermal
occupation number) is required to observe a negative weak value.}
\label{fig:nwetanth}
\end{figure}

The probability has been plotted as a function of detector
efficiency $\eta$ and the average thermal occupation number
$n_{th}$ in Fig. \ref{fig:nwetanth}. We see that for finite
thermal occupation number $n_{th}$ there is a lower bound on the
detector efficiency $\eta$ to see nonclassical behavior. This
bound vanishes when $n_{th} \rightarrow 0$.

\section{Conclusions}

In this work we have examined a very general form of weak
measurements focusing on the emergence of nonclassical features.
We have shown that the appearance of strange weak values is
equivalent to the existence of negative values for a generalized
Terletsky-Margenau-Hill distribution.

We have presented some feasible practical implementations of this
kind of measurement in the field of quantum optics focusing on the
weak measurement of photon number and energy. We have demonstrated
that negative weak values of energy contradict all classical models
of light, and that negative weak values of photon number contradict
a classical model where the energy is bounded from below by the
zero-point energy.

As a particular but striking enough example we have considered weak
measurements on coherent states. We have found that negative weak
values can be observed with a probability of 16 \% for kinetic
energy (or squared field quadrature), 8 \% for harmonic oscillator
energy and 50 \% for photon number.

We have analyzed the persistence of the effect under practical
experimental conditions by considering degrading imperfections
such as the presence of thermal fluctuations and the use of
inefficient detectors. All these results confirm the possibility
of a practical observation of nonclassical effects for states
previously considered as firm examples of classical behavior.

\appendix

\section{Imperfect detectors}\label{app:povm}

An imperfect detector may be represented by a Positive Operator
Valued Measure (POVM). To represent imperfect detection of an
observable $\hat{\phi}$, we study the class of diagonal POVM's
\begin{equation}\label{eq:diagPOVM}
    \hat{\Pi}_\phi = \int d\phi' \; \Pi_\phi(\phi')
    \mid \phi' \rangle \langle \phi' \mid,
\end{equation}
where $\mid \phi \rangle$ are eigenstates of $\hat{\phi}$. These
states are assumed to constitute a complete set,
\begin{equation}
    \int d\phi \mid \phi \rangle \langle \phi \mid = \mathrm{I}.
\end{equation}
The diagonal form is assumed because interference between
different detector states should not occur. We assume that
$\Pi_\phi(\phi')$ is a nonnegative function.

The POVM should provide a resolution of the identity operator
\begin{equation}
    \int d\phi \; \hat{\Pi}_\phi = \mathrm{I}.
\end{equation}
This implies that
\begin{equation}\label{eq:normalization}
    \int d\phi \; \Pi_\phi(\phi') = 1.
\end{equation}
Thus, $\Pi_\phi(\phi')$ should be a normalized distribution over
$\phi$. The probability distribution for the observable
$\hat{\phi}$, taking into account the imperfect detector
represented by $\Pi_\phi$, is
\begin{equation}\label{eq:probdistr}
    \mathrm{Tr} \left [ \hat{\Pi}_\phi \hat{\rho}
    \right ] = \int d\phi' \, \Pi_\phi(\phi') \langle \phi' \mid
    \hat{\rho} \mid \phi' \rangle.
\end{equation}
We require that the imperfect detector should give the same
expected reading as a perfect detector. This condition of
unbiasedness can be written as
\begin{equation}
    \int d\phi \; \phi \; \mathrm{Tr} \left [ \hat{\Pi}_\phi
    \hat{\rho} \right ] = \int d\phi \; \phi \; \langle \phi \mid
    \hat{\rho} \mid \phi \rangle,
\end{equation}
and it implies that
\begin{equation}\label{eq:unbiased}
    \phi' = \int d\phi \; \phi \; \Pi_\phi(\phi').
\end{equation}
Thus, unbiasedness is equivalent to requiring that the parameter
$\phi'$ should be the expectation value of the distribution
$\Pi_\phi$.

A typical diagonal POVM satisfying the properties
(\ref{eq:normalization}) and (\ref{eq:unbiased}) is the Gaussian
\begin{equation}\label{eq:gaussian}
    \Pi_\phi(\phi') = {1 \over {\sqrt{2 \pi \sigma^2}}} \;
    e^{-(\phi-\phi')^2/(2\sigma^2)}
\end{equation}
For this POVM, and for the subset of POVM's where
$\Pi_\phi(\phi')$ is a function of $(\phi-\phi')$ only, the
transformation (\ref{eq:probdistr}) is a convolution. However, in
the most general case the POVM does not depend on $(\phi-\phi')$
only. For example, the conditions (\ref{eq:normalization}) and
(\ref{eq:unbiased}) are also satisfied by all Gaussian POVM's of
the type (\ref{eq:gaussian}) even when the standard deviation
$\sigma$ is an arbitrary function of $\phi'$.

\section{Derivation of weak values}\label{app:weak}

Prior to the measurement interaction, we assume that the object is
in the state $\hat{\rho}_s$ and the pointer is in an arbitrary
mixed state $\hat{\rho}_a$. The total density operator prior to
the interaction has the product form
$\hat{\rho}_0=\hat{\rho}_s\otimes \hat{\rho}_a$.

After the interaction, the total density operator has evolved to
\begin{equation}
    \hat{\rho}_\epsilon = \hat{U}_\epsilon \hat{\rho}_0
    \hat{U}_\epsilon^\dag,
\end{equation}
where $\hat{U}_\epsilon$ is a unitary evolution operator. Since
the Hamiltonian does not depend explicitly on time, we can write
\begin{equation}
    \hat{U}_\epsilon = e^{-i \int \hat{H} dt} = e^{- i \epsilon
    \hat{\nu} \otimes \hat{P}}.
\end{equation}
We may expand this evolution operator to first order in $\epsilon$
as
\begin{equation}
    \hat{U}_\epsilon \approx 1 - i \epsilon \; \hat{\nu} \otimes
    \hat{P}.
\end{equation}
To first order in $\epsilon$, the density operator after the
interaction then can be written as
\begin{equation}\label{eq:firstorderdensity}
    \hat{\rho}_\epsilon = \hat{\rho}_0 + i \epsilon \;
    [\hat{\rho}_0,\hat{\nu} \otimes \hat{P} ] .
\end{equation}
After the interaction, we assume that imperfect measurements are
made of the pointer observable $\hat{Q}$ and the object observable
$\hat{\phi}$. Each observable is represented by a diagonal POVM
\begin{eqnarray}\label{eq:objectpovm}
    \hat{\Pi}_\phi &=& \int d \phi' \; \Pi_\phi(\phi') | \phi'
    \rangle \langle \phi' |, \\
    \hat{\Pi}_Q &=& \int dQ' \; \Pi_Q(Q') | Q' \rangle
    \langle Q' |.
\end{eqnarray}
Both $\Pi_\phi$ and $\Pi_Q$ are assumed to be classical
distributions. In this way, the detectors are assumed to be in a
statistical mixture of pure projector states. We also have assumed
that the observable $\phi$ has a continuous spectrum. There is no
loss of generality in this.

The joint probability density for $\phi$ and $Q$ after the
interaction is
\begin{equation}
    \rho_\epsilon (\phi,Q) = \mathrm{Tr} \, ( \hat{\Pi}_\phi \hat{\Pi}_Q
    \hat{\rho}_\epsilon),
\end{equation}
where the trace is taken over both ${\cal H}_s$ and ${\cal H}_a$.
Using the expression (\ref{eq:firstorderdensity}) for the first
order density operator, we find that
\begin{widetext}
\begin{eqnarray}
    \rho_\epsilon (\phi,Q) &=&
    \mathrm{Tr} (\hat{\Pi}_\phi \hat{\rho}_s) \mathrm{Tr} (\hat{\Pi}_Q
    \hat{\rho}_a) + i \epsilon \, \mathrm{Tr} \;
    (\hat{\Pi}_\phi \hat{\Pi}_Q [\hat{\rho}_0,\hat{\nu} \otimes \hat{P} ])
    \nonumber \\
    &=& \mathrm{Tr} (\hat{\Pi}_\phi \hat{\rho}_s) \mathrm{Tr} (\hat{\Pi}_Q
    \hat{\rho}_a) + i \epsilon \, \left [ \mathrm{Tr} (
    \hat{\Pi}_\phi \hat{\rho}_s \hat{\nu}  ) \mathrm{Tr}  (
    \hat{\Pi}_Q \hat{\rho}_a \hat{P} ) - \mathrm{Tr}  (
    \hat{\Pi}_\phi \hat{\nu} \hat{\rho}_s  ) \mathrm{Tr}  (
    \hat{\Pi}_Q \hat{P} \hat{\rho}_a  ) \right ].
\end{eqnarray}
\end{widetext}
We require that the current density of the pointer vanishes,
\begin{equation}\label{eq:zerocurrent}
    \langle Q \mid \left ( \hat{P} \hat{\rho}_a + \hat{\rho}_a
    \hat{P} \right ) \mid Q \rangle = 0.
\end{equation}
This implies that
\begin{equation}
    \mathrm{Tr} \; \left [ \hat{\Pi}_Q ( \hat{P}
    \hat{\rho}_a + \hat{\rho}_a \hat{P} ) \right ] = 0.
\end{equation}
Therefore we can write
\begin{eqnarray}
    \rho_\epsilon (\phi,Q) &=&
    \mathrm{Tr} (\hat{\Pi}_\phi \hat{\rho}_s) \mathrm{Tr} (\hat{\Pi}_Q
    \hat{\rho}_a) \nonumber \\ &+& i \epsilon \,
    \mathrm{Tr} \left [ \hat{\Pi}_\phi \left ( \hat{\rho}_s \hat{\nu}
     + \hat{\nu} \hat{\rho}_s \right ) \right ] \mathrm{Tr} ( \hat{\Pi}_Q
    \hat{\rho}_a \hat{P}).
\end{eqnarray}
The marginal distribution for the object observable $\phi$ after
the interaction is
\begin{equation}
    \rho_\epsilon(\phi) = \int dQ \; \rho_\epsilon (\phi,Q) .
\end{equation}
Due to the vanishing of the current density of the probe, it is
found that
\begin{equation}\label{eq:undisturbed}
    \rho_\epsilon(\phi) = \mathrm{Tr} (\hat{\Pi}_\phi
    \hat{\rho}_s).
\end{equation}
Therefore, the probability distribution for the postselection
observable $\phi$ is unaffected by the measurement interaction.
Note that we have considered an arbitrary postselection
measurement. This means in fact that the probability distribution
for every possible object observable is unaffected by the
measurement interaction.

The conditional probability density for the pointer position $Q$
given the postselection outcome $\phi$ is defined as
\begin{equation}
    \rho(Q \mid \phi) = {\rho_\epsilon (\phi,Q) \over
    \rho_\epsilon(\phi)}.
\end{equation}
We find that
\begin{equation}
    \rho(Q \mid \phi) =
    \mathrm{Tr} (\hat{\Pi}_Q \hat{\rho}_a) + 2 i \epsilon \,
    \mathrm{Re} (\nu_w) \mathrm{Tr} ( \hat{\Pi}_Q \hat{\rho}_a \hat{P}),
\end{equation}
where
\begin{equation}
    \nu_w(\phi) = {\mathrm{Tr} ( \hat{\Pi}_\phi \hat{\nu} \hat{\rho}_s ) \over
    \mathrm{Tr} (\hat{\Pi}_\phi \hat{\rho}_s)}
\end{equation}
is the weak value of $\hat{\nu}$ for an unsharp postselection
represented by the POVM $\hat{\Pi}_\phi$. Using Eq.
(\ref{eq:zerocurrent}) it can be shown that
\begin{equation}
    \langle Q \mid \hat{\rho}_a \hat{P} \mid Q \rangle = {i \over
    2} {\partial \over \partial Q} \langle Q \mid \hat{\rho}_a
    \mid Q \rangle.
\end{equation}
Hence we can write
\begin{eqnarray}
    \mathrm{Tr} \; ( \hat{\Pi}_Q \hat{\rho}_a \hat{P} ) &=& \int dQ''
    \int dQ' \Pi_Q(Q') \langle Q'' | Q' \rangle \langle Q^\prime
    | \hat{\rho}_a
    \hat{P} | Q'' \rangle \nonumber \\
    &=& \int dQ' \, \Pi_Q(Q') \, \langle Q' | \hat{\rho}_a \hat{P} | Q'
    \rangle \nonumber \\
    &=& {i \over 2} \int dQ' \, \Pi_Q(Q') \, {\partial \over \partial
    Q'} \langle Q' | \hat{\rho}_a | Q' \rangle \nonumber \\
    &=& - {i \over 2} \int dQ' \, {\partial \Pi_Q(Q') \over \partial Q'}
    \, \langle Q' | \hat{\rho}_a | Q' \rangle,
\end{eqnarray}
where we have assumed that $\langle Q' | \hat{\rho}_a | Q'
\rangle$ vanishes at the integration borders.

The conditional probability density for the pointer position $Q$
given the outcome $\phi$ of the postselection is
\begin{equation}\label{eq:condprob}
    \rho_\epsilon (Q \mid \phi) = \int dQ' \, \Pi_Q(Q' + \epsilon \mathrm{Re}\nu_w) \, \langle
    Q' | \hat{\rho}_a | Q' \rangle,
\end{equation}
where
\begin{equation}\label{eq:translated}
    \Pi_Q(Q' + \epsilon \mathrm{Re}\nu_w) = \left [ 1 + \epsilon
    \mathrm{Re}(\nu_w)  {\partial \over \partial Q'} \right ]
    \Pi_Q(Q').
\end{equation}
We introduce the expectation value of the pointer position
conditioned on the postselection outcome,
\begin{equation}
    E_\epsilon(Q \mid \phi) = \int dQ \; Q \; \rho_\epsilon (Q \mid
    \phi).
\end{equation}
By using Eqs. (\ref{eq:condprob}) and (\ref{eq:translated}) we
find that
\begin{eqnarray}
    E_\epsilon(Q \mid \phi) = E_0(Q \mid \phi) + \epsilon
    \mathrm{Re}(\nu_w) \nonumber \\ \times \int dQ \, Q \, \int dQ'
    {\partial \Pi_Q(Q') \over \partial Q'} \langle Q' \mid
    \hat{\rho}_a \mid Q' \rangle.
\end{eqnarray}
A reorganization of terms gives
\begin{eqnarray}
    E_\epsilon(Q \mid \phi) = E_0(Q \mid \phi) + \epsilon
    \mathrm{Re}(\nu_w) \nonumber \\ \times
    \int dQ' \, \langle Q' \mid \hat{\rho}_a \mid Q' \rangle
    {\partial \over \partial Q'} \int dQ  \, Q \, \Pi_Q(Q').
\end{eqnarray}
We assume that the POVM is unbiased, fulfilling condition
(\ref{eq:unbiased}). This, together with the normalization
property of the position distribution, implies that
\begin{equation}
    E_\epsilon(Q \mid \phi) = E_0(Q \mid \phi) + \epsilon
    \mathrm{Re}(\nu_w).
\end{equation}
This shows that the pointer observable $Q$ has been translated a
distance $\epsilon \mathrm{Re} (\nu_w)$, and this setup therefore
allows for a measurement of $\mathrm{Re} (\nu_w)$. The only
restriction on the auxiliary pointer system is that the current
density should vanish.

\end{document}